\documentstyle[aps,preprint]{revtex}

\tightenlines

\input epsf.sty

\begin{document}

\title{Pseudofractals and box counting algorithm}

\author{Andrzej Z. G\'orski\thanks{Address: Institute of Nuclear
Physics, Radzikowskiego 152, 31--342 Krak\'ow, Poland, e--mail:
Andrzej.Gorski@ifj.edu.pl}}
\address{Institute of Nuclear Physics, Cracow, Poland}

\date{\today}

\maketitle

\begin{abstract}
 We show that for sets with the Hausdorff--Besicovitch dimension
equal zero the box counting algorithm commonly used to
calculate Renyi exponents ($d_q$) can exhibit perfect scaling
suggesting non zero $d_q$'s.
Properties of these pathological sets ({\it pseudofractals}) are
investigated.
Numerical, as well as analytical estimates for $d_q$'s are obtained.
A simple indicator is given to distinguish pseudofractals
and fractals in practical applications of the box counting method.
Histograms made of pseudofractal sets are shown to
have Pareto tails.
\end{abstract}

\section{Introduction}

  The notion of fractal has been introduced in 70's by B. Mandelbrot
and soon it has become very fashionable.
In mathematical sense a set is called fractal (set) when its
Hausdorff--Besicovitch dimension ($d_{HB}$) is greater than its
topological dimension ($d_T$) \cite{Mandel77}.
Since fractality is strictly related to the physically important
self similarity (self affinity), scaling symmetries and the
renormalization group,
it is widely used in physics on all scales: ranging from particle
\cite{hepfrac}
to astrophysics \cite{astrofrac}, and in various other areas, like
solid
state physics \cite{Hofstadter} or econophysics \cite{econofractal}.

  However, in contrast to fractal sets constructed by mathematicians
like the famous triadic Cantor set (1883), for physically
interesting cases, the algorithms to construct corresponding data sets
are usually unknown and it is very difficult (or just impossible)
to calculate their Hausdorff--Besicovitch dimension, Renyi exponents
{\it etc.} in mathematically rigorous way.
  Instead, one considers a zero dimensional (finite number)
subset of data points and one applies a standard numerical
algorithm, like the box counting (BC) algorithm or its derivatives,
that gives the well known log--log plot.
A good linear fit is assumed to be equivalent to
the calculation of the corresponding fractal dimensions.
 Apart from the fact that the above fit is to some extent arbitrary
(see {\it e.g.} \cite{molteno}) and there is no good method
to calculate "error bars", it will be shown in
the following sections that even a perfect fit can be misleading.

 In fact, there are many different mathematical definitions of the
fractal (capacity) dimension that can give different
results when applied to the fractal set.
They have been originally introduced to physics to characterize
strange attractors of dynamical systems \cite{FOY}.
The BC capacity dimension is closest to the dimension introduced by
Kolmogorov \cite{Kolmogorov}.

 Here, we limit our discussion to the BC method that is
the basic paradigm for practical computation of the generalized Renyi
exponents, $d_q$, defined by \cite{Renyi,HPineq}
\begin{equation}
d_q = {1\over 1-q} \ \lim_{N\to\infty} { \ln \sum_i p_i^q(N)
\over \ln N} \equiv \lim_{N\to\infty} {\ln Y(N) \over \ln N}
\ ,
\label{dqdef}
\end{equation}
where $N$ is the total number of "boxes" (bins), $p_i$
is the part of the "mass" ({\it i.e.} fraction of all points)
contained in the $i$-th box.
Also, in this paper we deal with sets that are not fractals,
namely the discrete (point) sets.
For these sets one defines
\begin{equation}
p_i(N) = {n_i(N) \over n_{tot}} \ ,
\label{pidef}
\end{equation}
where $n_i(N)$ is the number of data points ("mass") in the $i$-th box
for a given subdivision (partition) $N$ and
$n_{tot}$ is the total number of data points ("mass") contained in all
boxes.
  In the case $q=0$ (capacity dimension) eq. (\ref{dqdef}) becomes
\begin{equation}
d_0 = \lim_{N\to\infty} { \ln M(N) \over \ln N } \ ,
\label{dq0def}
\end{equation}
where $M(N)$ denotes just the number of non empty boxes.
In this case the number of data points in
particular boxes is irrelevant and this singles out the value $q=0$.
This is the reason why the BC method gives a unique result
for $d_0$ (see Sec. II).
$d_q$ is determined from the log--log plot of $\log Y(N)$
{\it vs.} $\log N$ with $N = 2^0, \ldots, 2^k$,
usually with $k \simeq 10 \div 30$.

  Since in practical computations with the BC and
derivative methods one always deals with finite number of data points,
we limit our analysis to the discrete sets.
  In the following Section we obtain an analytic expression for $d_0$
in case of the set defined by\cite{Politi,AZG}
\begin{equation}
x_n = {1 \over n^a} \ , \quad n=1, 2, \ldots \ ,
\quad a > 0 \ .
\label{xndef}
\end{equation}
The same method is also applied for general discrete sets with an
accumulation point, as well as for divergent series.
 In Sec. III the BC algorithm is applied to calculate
the Renyi exponents with $q \ne 0$ for (\ref{xndef}).
The excellent scaling (linear fit) has been found in full agreement
with analytical estimates, in spite of the fact that the set
(\ref{xndef}) is not a fractal and has the Hausdorff--Besicovitch
dimension equal to zero.
Also, it is shown that the standard BC method leads to a violation of
the Hentschel--Procaccia inequality\cite{HPineq}.
 A modification of the standard BC method which preserves
the HP inequality is analyzed as well.
Our results can be generalized to sets with an arbitrary number of
accumulation points.
In Sec. IV it is shown that pseudofractals generate histograms with
fat tails, in contrast to fractals.
The final discussion is given in the last Section.

\section{Capacity dimension of pseudofractals}

 Clearly, the discrete and countable set (\ref{xndef}) is
not a fractal and it has zero Hausdorff--Besicovitch dimension.
However, as has been demonstrated using the dimension function
\cite{Politi} or by direct application of the BC method
\cite{AZG}, numerical computation must give the following
analytic result
\begin{equation}
d_0 = {1 \over 1 + a} \ .
\label{d0ofa}
\end{equation}

 As the method of analytical estimates of \cite{AZG} and its
generalization will be used throughout the paper, we now describe
it briefly. Assuming the unit size of the whole set, the size of a
single bin is $1/N$. Denoting by $N_{sngl}$ number of bins (and by
$n_{sngl}=N_{sngl}$  number of corresponding data points)
with one and only one data point inside, one can easily calculate
from (\ref{xndef})
\begin{equation}
n_{sngl} = N_{sngl} \sim N^{1\over 1+a}
\ .
\label{Nsngl}
\end{equation}
Since we have logarithm in (\ref{dqdef}) and (\ref{dq0def}),
and the limit $N\to\infty$, the constant pre-factor can be neglected.
 The remaining data points ($n_r$) are closer to each other than the
bin size. Hence, all those bins are not empty.
Number of such bins ($N_r < n_r$) is equal to the distance of the
point $x_{n_{sngl}}$ from the accumulation point ($x_\infty=0$)
divided by the bin size ($1/N$). This gives the estimate
\begin{equation}
N_r \sim N^{1\over a} \ ,
\end{equation}
and in the limit $N\to\infty$
\begin{equation}
M(N) \sim N_{sngl} + N_r \sim N^{1\over 1+a} + N^{1\over a}
\sim N^{1\over 1+a}
\ ,
\label{MNest}
\end{equation}
that implies the result (\ref{d0ofa}).

 From the above proof it is clear that the exponent $d_0$ depends on
the rate of change of the distances between neighboring points
("level spacing") with respect to the length of the whole interval or,
in other words, on the speed of convergence of data points to the
accumulation point.
This enables us to generalize the above result.
 Also, one can consider divergent sets
($x_n\to\infty$ for $n\to\infty$) by rescaling them to the unit
interval. To this end, we define the convergence rate $\Delta x(n)$
by
\begin{equation}
\Delta x(n) = \left\{
              \begin{array}
              [l]{ll}
              {\vert x_n - x_{n+1} \vert /
              \vert x_1 - x_\infty \vert}
              \quad & {\rm for} \ \vert x_\infty\vert < \infty \ ,
              \\
              {\vert x_n - x_{n+1} \vert /
              \vert x_n \vert}
              \quad & {\rm for} \ \vert x_\infty\vert = \infty \ .
              \end{array}
              \right.
\label{Deltaxdef}
\end{equation}
 This gives the following general formula for the exponent $d_0$
\begin{equation}
d_0 = \min \left\{ \lim_{n\to\infty} { -\ln n \over \ln
\Delta x(n)} , \ 1 \right\}
\ .
\label{d0formula}
\end{equation}
 In particular for (\ref{xndef}), one gets
$\Delta x(n) \sim 1/ n^a$ that leads to formula (\ref{d0ofa}).
 For slowly converging series, like $1/\ln n$, one has
$d_0 = 1$, while for strong convergence ({\it e.g.}
$x_n = e^{-an}$) there is $d_0 = 0$.
On the other hand for all diverging series (like $n^a$,
$e^{+an}$ or $\ln n$) there is always $d_0 = 1$.
Intuitively, slowly converging series look like uniformly
distributed, while those exponentially converging look like
concentrated at the accumulation point (zero dimensional).
Hence, from this point of view, the series with inverse power
asymptotic are the only non trivial ones.

\begin{figure}[bht]
\centering
\epsfxsize=8.5truecm
\mbox{
\epsfbox{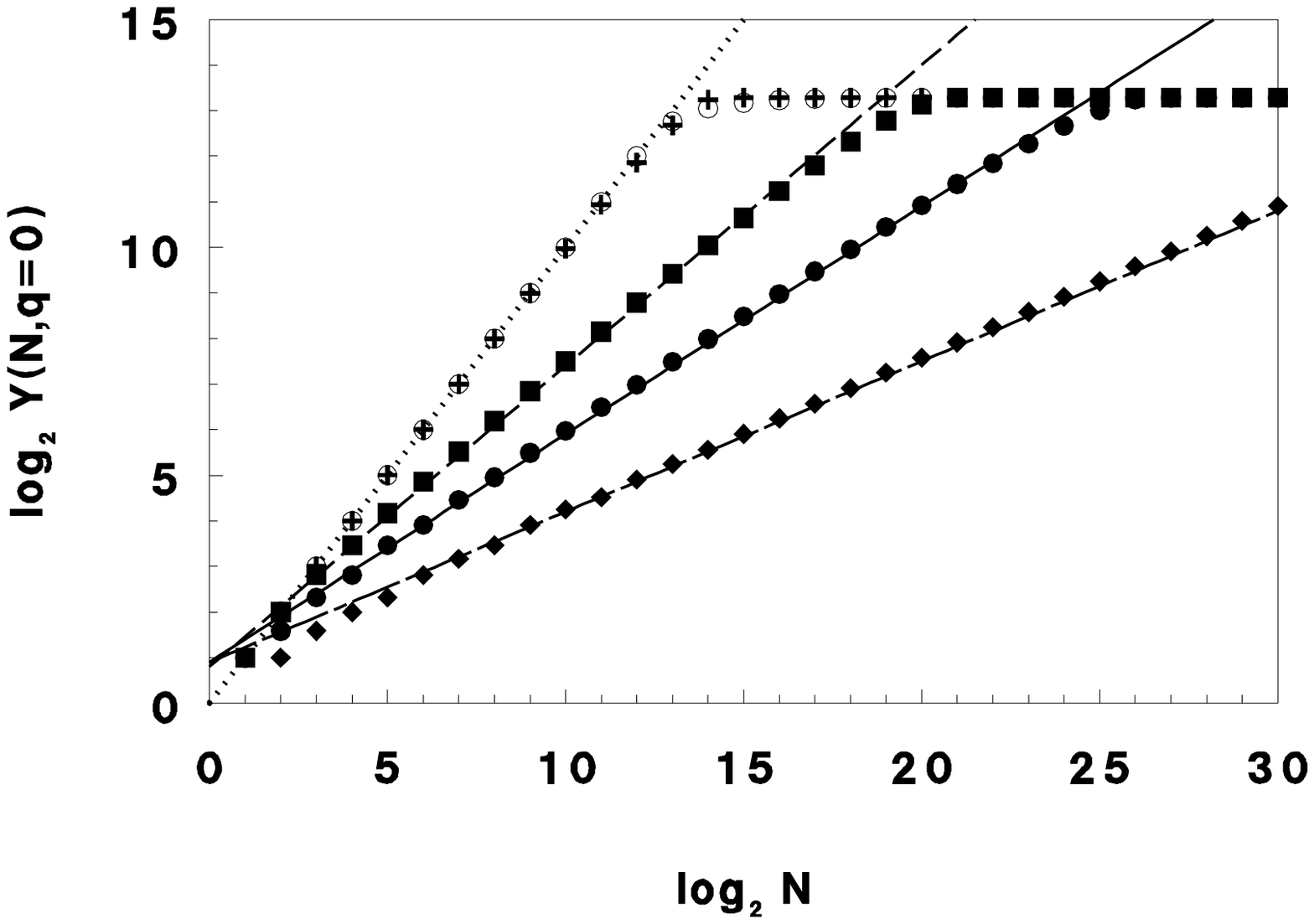}
}
\caption{Log--log plots and analytical predictions (lines)
for $d_0$ with: $x_n = 1/n$ (full circles and solid line),
$1/n^{1/2}$
(squares and dashed line), $1/n^2$ (diamonds and dashed-dotted line),
$n^{1/2}$
and $n^2$ (crosses and circles with one dotted line for
both).
}
\label{fig:Fig1}
\end{figure}

 The above results can be verified numerically by applying the BC
method. The results are displayed in Fig.~\ref{fig:Fig1},
where straight lines correspond to the theoretical predictions.
Actually, already for $10^4$ data points one can see an excellent
linear
scaling in the log--log plot throughout more than dozen of binary
orders of magnitude --- well above of what is usually demanded
in practical applications. In addition, the results
are in perfect agreement with formula (\ref{d0formula}):
$d_0 = 0.50$, $0.66$ and $0.33$ for $a = 1$, $0.5$ and $2$,
respectively, while for divergent series, $\sqrt{n}$ and $n^2$
(crosses and circles), one obtains $d_0 = 1.0$.

\section{Pseudofractals and generalized Renyi exponents}

 For $q\ne0$ analytical estimates are ambiguous as we have to
deal with the double limit:
$\lim_{N\to\infty} \lim_{n_{tot}\to\infty}$, because the
probabilities
($p_i = p_i(n_i, n_{tot}, N)$) do depend on both, $N$ and $n_{tot}$.
Equivalently, the measure (\ref{pidef}) is not well defined.
 In standard applications of the BC method one has fixed number
of data points ($n_{tot}=$ const.) and the large $N$ limit is
being estimated.
In this case, for $q\le0$ one can estimate the sum in (\ref{dqdef})
as in the derivation of (\ref{MNest}), by taking partial sum with bins
containing only one data point. Namely,
\begin{eqnarray}
{1\over 1-q} \ln \sum_{i=1}^{N_{sngl}} p_i^q &&=
{1\over 1 - q} \ \ln \left[ N^{1\over 1+a}
\left( {1\over n_{tot}} \right)^q \right] = \nonumber\\
&&= {\rm const} + {1\over 1-q} {1\over 1+a} \ln N
\ .
\nonumber
\end{eqnarray}
The upper limit can be estimated assuming equal number of data points
in remaining bins ($N_r \sim N^{1/a}$).
One should remember, that due to the limited number of data points
the number of bins cannot be too large: $1 \ll N < n_{tot}^{1+a}$.
For finer partitions we reach the saturation point ---
there is a constant number of non empty bins with exactly one data
point inside, that corresponds to the value of
$\log Y_{max} = \log n_{tot}$
(see Figs.~\ref{fig:Fig1} and \ref{fig:Fig2}(A), where
$\log_2 Y_{max} = \log_2 10^4 \simeq 13.3$).
 Finally, we obtain an analytical estimate for large $N$
\begin{equation}
d_q = {1\over 1-q} \ {1\over 1+a}
\qquad (q\le 0)
\ .
\label{estimateA}
\end{equation}
 For $q>0$ estimates become more complicated, as truncation of the
sum can make it smaller than one that causes the change of sign
of the logarithm. However, for large $q$ one gets fast convergence
$d_q\to 1$ ($q\to +\infty$).
 Again, as it is clear from Fig.~\ref{fig:Fig2}(A), we obtain
very good linear fits throughout about ten binary orders of magnitude
that is usually interpreted as a sign of fractality and
excellent agreement with the theoretical estimate.

 It has been proven that for fractal sets Renyi exponents
$d_q$ the HP inequality holds \cite{HPineq}
\begin{equation}
d_q \le d_{q^\prime} \quad {\rm for } \ q > q^\prime
\ .
\label{HPinequality}
\end{equation}
 However, in our case the calculated scaling exponents
apparently violate (\ref{HPinequality}) as can be seen form
(\ref{estimateA}) and from
Figs.~\ref{fig:Fig2}(A)~and~\ref{fig:Fig3} (full circles and the
dashed line).

 Let us notice that calculating $d_q$'s analytically, for well defined
fractal sets, like the triadic Cantor set, the resolution for counting
data points increases when the bin number is increasing.
And the resolution at a given step (for a given partition)
is equal to the bin size (smallest void intervals are of the bin
size). This is in contrast to the standard version of the BC method,
where the data set is fixed during the whole procedure.
 Now, let us modify the BC method by taking into account
for a given partition only those points that are separated
from each other by at least the (current) bin size ({\it i.e.}
the bin size fixes the resolution).
This makes the computation more involved and time consuming but,
in effect, one can recover the HP inequality.

 For the modified BC method, in the way similar as for the estimate
(\ref{estimateA}), one obtains the following analytical formula
\begin{equation}
d_q = {1\over 1-q} \ \left[ {1\over 1+a} - {q\over a} \right]
\qquad (q\le 0)
\ .
\label{estimateB}
\end{equation}
In addition, one has $d_q\to 0$ for ($q\to+\infty$).
Clearly (\ref{estimateB}) satisfies
(\ref{HPinequality}).
This result can be validated numerically, as is displayed
in Fig.~\ref{fig:Fig3} (full squares and the dotted line
for eq. (\ref{estimateB})). Again, we have a
very good scaling and linear fit.
 For positive $q$ this method gives $d_q$ that tends to zero quite
fast (while it reaches one, a bit slower, for the standard BC
algorithm).
 Hence, for pseudofractals the standard and modified BC algorithms
give different results due to ambiguity mentioned earlier.
However, in both cases a good scaling and linear fit is obtained.

 Our conclusions remain unchanged for sets with an arbitrary number of
accumulation points. In particular, the union of two sets with
scaling exponents $d_0^{(1)}, d_0^{(2)}$ in the large number
of bins ($N$) limit gives
\begin{eqnarray}
d_0 \ &&= \ {1\over \ln N} \ \ln \left[ N^{d_0^{(1)}} + N^{d_0^{(1)}}
\right] \ = \ \max \left\{ d_0^{(1)}, d_0^{(2)} \right\} +
\nonumber \\
&&+ {1\over \ln N} {1\over N^{\vert d_0^{(1)} - d_0^{(2)} \vert} }
\to \max \left\{ d_0^{(1)}, d_0^{(2)} \right\}
\ .
\label{d0sum}
\end{eqnarray}
Like for regular fractals, the scaling exponent of the
union is equal to the maximal exponent of the two sets.
However, notice that convergence to this result (in the large $N$
limit) is slowest (logarithmic) for $ d_0^{(1)} =  d_0^{(2)}$.
As the number of non empty bins is greater than for a single set,
points in the log--log plot will be higher for not too large $N$.
Hence, the whole plot will be a bit less steep and for not large
enough data sets this can lead to lower estimates of the Renyi
exponents and worse linear fit. Details of this effect depend on the
particular distribution of both sets in the embedding interval
and have been verified numerically.
For sets not large enough the scaling can be completely lost.

\section{Pseudofractals and fat tails}

 One is often interested in probability distribution for a large
series of data. In particular, in recent years there has been a great
interest in the so--called Pareto or fat tails \cite{Stanley},
where histograms built out of the data have inverse power law tails
\begin{equation}
P(x) \ \sim \ 1/x^\beta
\ ,
\label{fattail}
\end{equation}
$P(x)$ being the probability distribution.
Here we show that non trivial ($0<d_0 < 1$)
pseudofractals do have this property.

 As for the histogram the time ordering of the data points can be
neglected, let us consider for simplicity a monotonic series
$\{ x_n: \ x_{n+1} \le x_n  \}$.
To satisfy (\ref{fattail}) the number of data points ($\Delta n$)
in the interval $[ x_{n+\Delta n}, x_n ]$ must be
\[
\Delta n = \int_{x_{n+\Delta n}}^{x_n} P(x) \ dx =
{C\over \beta} \ \left[
{1\over x_{n+\Delta n}^{\beta-1}} - {1\over x_n^{\beta-1}}
\right] \ ,
\]
where $C>0$ is a normalization constant.
Substituting $C_1 = \beta/C > 0$   and
$f(n) \equiv 1/x_n^{\beta-1}$
this yields the following simple linear first order difference
equation
\[
C_1 \ \Delta n \ = \  f(n+\Delta n) - f(n)
\ ,
\]
with the general solution $f(n) = C_1 \ n + C_2$
or, equivalently
\[
x_n \ = \ {1\over \left[ C_1 \ n +
C_2 \right]^{1\over \beta-1}}
\ .
\]

 For tails ($n \gg C_2/C_1$ but still far from the accumulation point)
the constant $C_2$ can be neglected and finally we have the asymptotic
behaviour
\begin{equation}
x_n \ \sim \ { 1\over n^{1\over \beta - 1} } \ \equiv \
{1\over n^a}
\ .
\label{tailas}
\end{equation}
Hence, the tail exponent $\beta$ can be expressed in terms of
$a$ or $d_0$ as
\begin{equation}
\beta \ = \ {a + 1 \over a} \ = \
{1 \over 1 - d_0}
\ .
\label{beta}
\end{equation}
The above formula displays simple relation between the
psudofractal's
parameter $a$, the tail index $\beta$ and the box counting exponent
$d_0$.

\section{Conclusions}

\begin{figure}[bht]
\centering
\epsfxsize=8.5truecm
\mbox{
\epsfbox{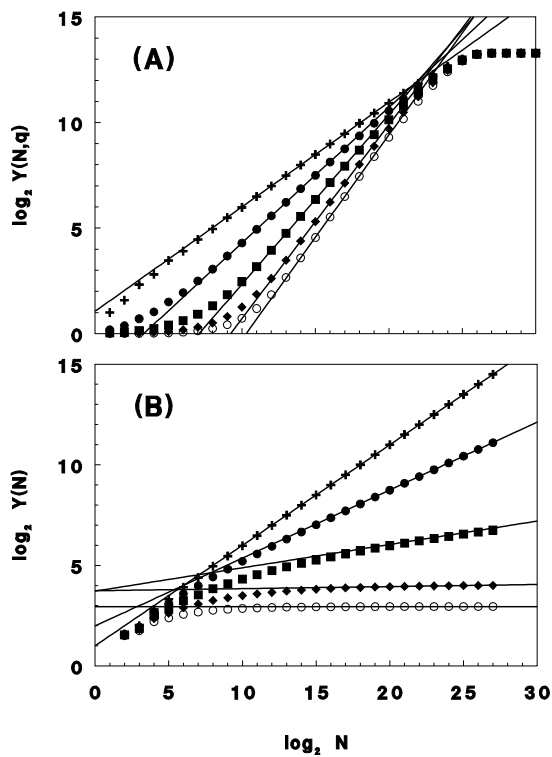}
}
\caption{Log--log plots for the harmonic series ($x_n = 1/n$)
with $q = 0$ (crosses), $0.25$ (full circles), $0.5$ (squares),
$0.75$ (diamonds) and $1$ (circles) with corresponding linear fits
(solid lines). The upper panel (A) is for $10^4$ data points
($n_{tot}=$const.)
and the standard BC method. The lower panel (B) is for the modified
BC algorithm.}
\label{fig:Fig2}
\end{figure}

\begin{figure}[bht]
\centering
\epsfxsize=8.5truecm
\mbox{
\epsfbox{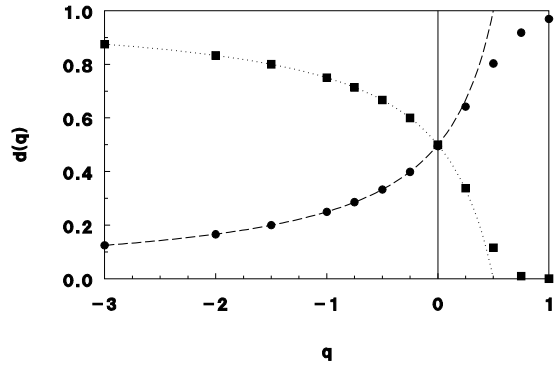}
}
\caption{$d(q)\equiv d_q$ computed for the harmonic series ($x_n =
1/n$) with $10^4$ data points for the BC (full circles)
and modified BC (full squares) methods.
Dashed and dotted lines represent analytic estimates
(11) and (13), respectively.}
\label{fig:Fig3}
\end{figure}

 In this paper, we investigate general sets with accumulation points,
that are not fractals, thought they display
fractal like scaling behaviour.
 The scaling exponent $d_0$ (eq. (\ref{dq0def})) as obtained by the BC
method is given by (\ref{d0formula}).
 Furthermore, we have found the analytical formula for $d_q$
(for $q\le0$) for the inverse power series as
given by the standard BC algorithm (eq. (\ref{estimateA})),
that perfectly fits to numerical results
(Fig.~\ref{fig:Fig2}(A)).
Obtained exponents violate the HP inequality,
that can be viewed as an indicator of the pseudofractal behaviour.

 Similar results are obtained for the modified BC algorithm
(where the number of data points taken into account is increasing
with the increased resolution), but in this case the HP inequality
is preserved (see eq. (\ref{estimateB}) and Fig.\ref{fig:Fig2}(B)).
Hence, the two schemes give different $d_q$'s for pseudofractal
sets.
 Our results remain valid for sets with arbitrary number of
accumulation points, where the overall scaling exponent is equal to
the maximal exponent of constituent sets. Also, in this case
one can observe worsening of the linear fit.

 In general, from the point of view of the fractal properties and the
BC methods there are four types of sets:

\noindent{\it (i) mathematical fractals} -- sets that are
well defined and their fractal properties can be rigorously proven
({\it i.e.} without numerical approximations), like the triadic
Cantor set.

\noindent{\it (ii) physical fractals} -- finite sets that are
(computer) representations of mathematical fractals.
 In this case one gets good scaling and linear fit with the BC method,
HP inequality holds and both BC and modified BC method
(described in Sec. III) give the same results.

\noindent{\it (iii) pseudofractals} -- finite sets that are not
finite representations of mathematical fractals, though
they show good scaling and linear fit with the BC method.
The resulting exponents violate the HP inequality and the BC and
modified BC algorithm give different values for $d_q$'s.
The general formula for $d_0$ in case when $x_n$ asymptotic
is known is given by (\ref{d0formula}).

\noindent{\it (iv)} non fractals, {\it i.e.} sets for which
the BC algorithm does not exhibit any scaling.

 The sets of types {\it (i)} and {\it (iv)} can be easily
distinguished. However, it is quite non trivial to distinguish between
sets of type {\it (ii)} and type {\it (iii)}. Here,
one cannot apply the rigorous mathematical machinery as the
whole set is usually unknown. In these cases
numerical methods lead to nice scaling making them impossible
to tell apart.
 In this context, violation of the HP inequality appears to be a
simple and useful indicator, in addition to different results
obtained by the standard and modified BC algorithms.

 Different classes of non trivial ($0<d_0 < 1$) pseudofractals
have scaling properties equivalent to the series
$\{ x_n = 1/n^a \}$.
In particular, for $q>0$ the BC method gives $d_q$ close
to the embedding dimension while for the modified BC algorithm
$d_q$ approaches zero.
For $q\le 0$ analytical formulae for $d_q$ are given by
(\ref{estimateA}) and (\ref{estimateB}), respectively.

 Finally, as shown by (\ref{beta}), the parameter $a$ of the
pseudofractal series is simply related to the tail index
$\beta$, as well as to the box counting Renyi exponent $d_0$.
This means that histograms made of non trivial pseudofractal sets
have Pareto (fat) tails.
This relation is another signal of possible pseudofractality.

\end{document}